\documentstyle[prl,aps,epsf,epsfig,floats]{revtex}
\begin{document}
\twocolumn[\hsize\textwidth\columnwidth\hsize\csname@twocolumnfalse\endcsname
%
\title{\bf Electronic Structure of Sr$_2$FeMoO$_6$} 

\author{D.D. Sarma\cite{jnc}, Priya Mahadevan, T. Saha-Dasgupta\cite{snb}, 
Sugata Ray and Ashwani Kumar}

\address{Solid State and Structural Chemistry Unit, Indian Institute
of Science, Bangalore 560012, India}

\maketitle

\begin{abstract}
We have analysed the unusual electronic structure of Sr$_2$FeMoO$_6$
combining {\it ab-initio} and model Hamiltonian approaches. Our results 
indicate that there are strong enhancements of the intraatomic exchange strength
at the Mo site as well as the antiferromagnetic coupling strength between Fe and
Mo sites. We discuss the possibility of a negative effective Coulomb 
correlation strength ($U_{eff}$) at the Mo 
site due to these renormalised interaction strengths.
\vspace{0.2in}
\end{abstract}
\pacs{
71.45.Gm,75.10.-b,71.20.-b,79.60.Bm
}

]
\narrowtext

Colossal magnetoresistance in transition metal oxides \cite {Helm} has 
attracted a great deal of attention in recent times \cite{refs} owing to 
their technological as well as fundamental importance. The systems based on 
doped manganites have been most exhaustively studied and their behaviors
have been broadly understood in terms of the double-exchange mechanism
\cite{Zene,Hase}, possibly modified by the presence of various kinds of
polarons \cite{polarons}. 
Recently, an ordered double perovskite oxide without
any manganese, Sr$_2$FeMoO$_6$, has been shown \cite{Koba} to have CMR at a 
low-field even at  room temperature. 
This spectacular MR in Sr$_2$FeMoO$_6$ is intimately 
connected with a large magnetic $T_c$ ($\approx$ 450 K) of this compound.
The essential physics of Sr$_2$FeMoO$_6$ is 
believed to be very similar to that of the manganites.
However this apparent similarity 
between the manganites and Sr$_2$FeMoO$_6$ is surprising, if not totally
unexpected.

Mo is not a strongly correlated system and consequently, a magnetic 
moment at the Mo site is a rarity, in contrast to Mn.
In the ordered double perovskite Sr$_2$FeMoO$_6$, the 
transition metal sublattice sites are occupied alternately by Fe and Mo ions. 
Experimental evidence of ferrimagnetic order as well as band 
structure results establish an {\em antiferromagnetic} coupling 
between Mo and Fe atoms, in 
contrast to the {\em ferromagnetic} coupling in manganites. 
The large magnetic transition temperature in Sr$_2$FeMoO$_6$, however, 
points to a large interatomic exchange coupling 
strength, $J$, between Fe and Mo, comparable to that between the Mn-Mn pairs in the manganites, inspite of the normally expected nonmagnetic nature of Mo. Mo is usually nonmagnetic as the intraatomic exchange 
strength, $I$, within the Mo 4$d$ manifold is small, typically 0.1-0.2~eV, nearly an order of magnitude smaller than that in the 3$d$ transition metal systems \cite{ours}. 
Additionally, the Mo 4$d$ bandwidth tends to be substantially larger 
compared to the exchange splitting.
However, the magnetic structure
of this compound requires a strong spin-splitting 
of the delocalized bands, leading to  nearly complete spin polarization. 
This issue 
is moot in the present context, since the spectacular properties of this 
system - a high temperature {\em and} low-field MR - are critically 
dependent on the large $T_c$ and the half-metallic ferromagnetic state. 
We investigate the origin of these apparent anomalies by combining 
state-of-the-art {\it ab initio} methods as well as realistic multiband 
many-body models for electronic structure calculations. Our results clearly 
establish an unusual renormalization of the intraatomic exchange strength at 
the Mo sites, arising from Fe-Mo interactions; this causes the {\em 
effective} $I$ to be much larger than the expected atomic value. We further 
evaluate $J$ to show that it is indeed large, owing to the renormalization
of $I$. Finally, we suggest the novel possibility of a negative effective 
Coulomb interaction strength, $U_{eff}$ (= $U$ - $I$), for the Mo sites  due to the enhancement of $I$, with some preliminary experimental
evidence for the same. Our results underline the unusual and unique 
aspects of the 
electronic structure of Sr$_2$FeMoO$_6$, that
have not been appreciated so far.  

We have computed the band dispersion of geometry optimised \cite{Koba} 
Sr$_2$FeMoO$_6$ within the framework of 
linear muffin-tin orbital (LMTO) method \cite{lmto} using the generalized gradient 
approximation (GGA) for the exchange-correlation part.
Figs. 1(a) and 1(b) show the orbital projected band dispersions 
of Sr$_2$FeMoO$_6$
for the majority and minority spin channels; here
the fatness of the bands in each panel is the weight of the indicated 
orbital in the wavefunction. 
The minority {\it down} spin bands
cross the Fermi level, while the majority {\it up} spin bands exhibit
a bandgap of $\sim$ 0.8 eV. The bands below -2~eV are predominantly of
oxygen character while bands crossing the Fermi level and ranging from 
-2~eV to about
2~eV have significant mixing between the Fe-$d$ and Mo-$d$ 
characters with some small admixture
of oxygen $p$ states. The presence of approximate cubic symmetry of the
octahedral co-ordination of the oxygen atoms around the transition metal sites,
results in a splitting of the $d$ levels into $t_{2g}$ 
and $e_g$ orbitals. For the minority
spin channel, the Fe $t_{2g}$ and Mo $t_{2g}$ bands 
are partially filled while Fe $e_g$ 
and Mo $e_g$ bands remain empty. The occupied part of the bands near the Fermi level 
in majority spin channel [see Fig. 1(a)] are mainly composed of Fe $d$ character
which hybridizes with the oxygen $p$ states. The narrow bands lying immediately above the
Fermi level spanning an energy range of about 0.7~eV to 1.2~eV are predominantly from Mo $t_{2g}$ contribution,
while the Mo $e_g$ bands are further high
up in energy. 
In order to estimate the electronic interactions strengths, 
we fitted the {\it ab initio} band 
dispersions, shown in Fig. 1, in terms of a tight-binding model containing 
$d$ orbitals at the Fe and the Mo sites and the $p$ orbitals at the 
oxygen sites.
Such methods have been successful in 
obtaining realistic estimates of the interaction strengths including 
that for the intraatomic exchange interactions \cite{ours,tbfita}. 
This analysis yields a value of the intraatomic exchange
splitting strength at the Mo site to be 0.13~eV, estimated from the spin
polarization of the Mo site energies; this estimate is consistent with 
our expectation of a small $I$ at the Mo sites, but is in 
apparent contradiction to the enhanced $T_c$ and strong polarization of 
the bands in Sr$_2$FeMoO$_6$. It is intriguing at this stage to note that 
the up-spin bands related to the Mo $d$ states appear in
the energy window centred at $\sim$ 1~eV, while 
the down-spin states appear near the Fermi level with a
larger bandwidth suggesting 
a spin-splitting of the Mo $d$ band, 
considerably larger than the bare $I$. This clearly indicates that the bare
$I$ at the Mo site must be strongly renormalised giving rise to an enhanced 
$I_{eff}$ leading to the pronounced spin polarization of 
the Mo $d$ bands. In order to understand this renormalisation process, 
as well
as to estimate the effective $I$ in contrast to the bare $I$ at the Mo sites, 
it is necessary to integrate out the other degrees of freedom, such as the 
Fe $d$ states, from the {\it ab initio} results, so that the few orbital 
description retains the effect of renormalisations as a change in the 
{\em effective} interaction strengths.

Very recently, the third generation LMTO
method \cite{lmto3} has been proved to be a powerful tool for deriving 
few orbital, orthogonal
tight-binding Hamiltonians starting from the all orbital Hamiltonian 
representation used for the local density approximation 
 or GGA self-consistency procedure, 
by integrating out other degrees of freedom.

In order to estimate the $I_{eff}$ at the Mo sites, responsible for the 
large spin polarization of the bands, 
we have used this technique to derive the
effective Mo $t_{2g}$ Hamiltonian from the all orbital Hamiltonian.
Figs. 2(a) and 2(b) show 
the energy
bands in {\it up} and {\it down} spin channels respectively, obtained by 
diagonalising the 3x3 Hamiltonian defined in the
effective Mo $t_{2g}$ basis. It is clear that this very powerful approach
captures the dispersions of the Mo $t_{2g}$
orbitals very effectively. The spin-splitting of the on-site
Mo $t_{2g}$ energies can be directly read off from the diagonal terms
of the real space representation of the Hamiltonian matrix 
(obtained by Fourier transformation of the downfolded Hamiltonian
in k-space $H(k) \rightarrow H(R))$
and was found to be 0.8 eV. For fully polarized Mo $t_{2g}$ 
electrons this
is then the {\it ab-initio} estimate of $I_{eff}$, while taking 
into account the LMTO estimate
for the magnetization (N$_{\uparrow}$-N$_{\downarrow}$=0.34 $\mu_B$) increases the 
$I_{eff}$ value to $\sim$ 2.3 eV.
This spectacular enhancement of the $I_{eff}$ compared to the bare $I$ of Mo
is in fact easy to rationalize in terms of the details of the electronic  
interactions in such a double perovskite system. Within the band structure 
calculations performed for the nonmagnetic ground state we find the Mo 4$d$ states 
to be located at higher energies compared to  the Fe 3$d$ states. By fitting the nonmagnetic band structure
to a tight-binding model, this  energy separation is found to be about
1.4~eV.  Considering 
the ferrimagnetic arrangement of the Fe and Mo sites, 
the orbital energies are as shown in the left panel of 
Fig.~3 in the absence of any hopping interaction.
The Fe$^{3+}$ site has a larger exchange splitting compared
to the crystal field splitting,  while the situation is reversed for the 
Mo site. In the presence of hopping interactions, there is 
finite coupling between states of the same symmetry at the 
Fe and the Mo sites, leading to  perturbations of the bare energy levels. 
It is 
then easily seen that the Mo $t_{{2g}_{\uparrow}}$ state will be pushed up 
and the Mo $t_{{2g}_{\downarrow}}$ state will be pushed further down by 
hybridization with the corresponding Fe states, as shown in the figure. 
These opposite movements of the Mo up- and down-states increase
the energy separation between these two states, thereby substantially increasing
the effective exchange splitting at the Mo site. Thus, the renormalisation of 
$I_{eff}$ at the Mo site is driven by the large $I$ at the Fe site and the 
substantial hopping interaction coupling the two sites. It is interesting to note,
and also follows from the arguments given above, that the antiferromagnetic
coupling between Fe and Mo is crucial for the enhanced 
$I_{eff}$ and consequently, the $T_c$. A 
ferromagnetic arrangement on the other hand would result in a reduced $T_c$.

 The above analysis as well as the estimate of $I_{eff}$ 
are based on effectively
single-particle theories. In order to verify that the 
present conclusions are not seriously affected by 
well-known limitations of such theories, we have 
also used many-body calculations to estimate $I_{eff}$. 
In order to obtain a realistic description of the specific system, we fixed 
the hopping interactions between the various orbitals to those estimated 
from the tight-binding fit of the full {\it ab initio} calculations. 
Furthermore, we performed Hartree-Fock calculations 
for the multiband model for the lattice involving Fe and 
Mo $d$ and oxygen $p$ states and reproduced the half-metallic 
ferromagnetic state in order to obtain appropriate estimates for the
charge transfer energies, intraatomic Coulomb and bare exchange strengths.
The parameters used are 0.65,  
1.34, -2.33, and 1.47 eV for oxygen $pp\sigma$, oxygen-Mo  
$pd\pi$, oxygen-Fe $pd\sigma$ and $pd\pi$ hopping strengths; $U_{Fe}$, 
$I_{Fe}$, $U_{Mo}$, $I_{Mo}$, $\Delta_{Fe}$, and $\Delta_{Mo}$ were estimated
to be 5.0, 0.8, 1.0, 0.2, 2.8 and 4.0 eV, respectively. Additional
crystal field splittings between the $t_{2g}$ and $e_g$ orbitals at the Fe 
site and Mo site were introduced.  

For calculation of $I_{eff}$, we consider a multiband
Hubbard-like Hamiltonian for the finite cluster 
Fe-O-Mo-O-Fe. The Hamiltonian includes on-site energies, hopping 
interactions and the on-site Coulomb interactions at Fe and Mo sites.
The full Hilbert space for even such a small 
cluster has a dimension of $\sim$ 2x10$^9$ ; 
since such a large calculation is not tractable, we included the hopping 
interactions connecting only the transition metal $t_{2g}$ orbitals to the 
oxygen $p_{\pi}$ orbitals. This is a reasonable approximation, as the 
relevant charge dynamics involves only these orbitals with the $e_g$ orbitals participating in
Coulomb interactions. Moreover, we restrict ourselves 
to the $S_z$=9/2 subspace in conformity with the experimentally observed 
antiferromagnetic arrangement of the transition metal moments. 
The renormalized $I_{eff}$ can be estimated 
from this cluster 
by calculating $\delta E/\delta n_{\uparrow} - \delta E/\delta n_{\downarrow}$, 
where $E$ is the total energy and $n_{\uparrow}$ and $n_{\downarrow}$ are 
the corresponding Mo occupancies. 
For the above set of parameter values, the 
$I_{eff}$ is calculated to be 0.95 eV
with integral charge ({\it i.e.} {$\delta$}n=1) fluctuations, 
exhibiting a large enhancement from the 
bare $I$ of 0.2 eV used in the calculation and in very good agreement with 
the estimate obtained from the {\it ab initio} results.
Since this estimate depends on the choice of the parameter strengths,
we have calculated $I_{eff}$ with reasonable variations of the
input parameters; we
find that
$I_{eff}$ is always between 0.8 and 1.5 eV.

In order to calculate the $J$, coupling the Fe and the 
Mo sites antiferromagnetically, we perform Hartree-Fock calculations with 
the transition metal $d$-oxygen $p$ multiband model with the parameter values
specified earlier. We considered different 
magnetic ground states corresponding to different values of 
the spin-density wave vector $q$. Small
rotations of the spin moments about the ground state gave us 
$J(q)$ \cite{eschrig}.
We then mapped the results on to a nearest neighbor Heisenberg model
and obtained a value of 18 meV for the coupling between Fe and Mo. A
similar analysis \cite{lamn} for the 
half-metallic 30$\%$ hole-doped LaMnO$_3$ gave us 12.5 meV.
Most importantly,
we find that the $J$ in Sr$_2$FeMoO$_6$ is in fact somewhat larger than that 
in the manganites, explaining the enhanced magnetic transition temperature
in the ordered double perovskite system compared to the hole doped manganites
($<$ 300~K). This enhanced value of $J$ is of 
course intimately connected with the strong spin-polarisation of the Mo $d$
states via the renormalised $I_{eff}$ at the Mo sites. 

It is to be noted that the exchange strength $I_{eff}$ at the Mo site 
is comparable or larger than the expected
value of the direct Coulomb interaction strength, $U$, at the Mo site. This 
leads to the intriguing situation that the effective electron-electron repulsion
strength, $U_{eff}$ = $U$ - $I_{eff}$, for Mo  may become negative. 
If this were indeed to be true, there should be some 
measurable consequences in terms of the electronic properties. In order to 
verify this, we have performed core level x-ray photoemission study of the 
Mo 3$d$ core level in Sr$_2$FeMoO$_6$. The core level spectrum is shown in Fig. 4.
For comparison, we also show the Mo 3$d$ spectra from two reference materials, 
Mo metal and MoO$_3$ in the insets A and B. The insets clearly show that the core 
level spectra 
from reference Mo systems contain only a doublet feature arising
from the spin-orbit splitting of the core level into 3$d_{5/2}$ and 3$d_{3/2}$
states. In sharp contrast, the Mo 3$d$ spectrum from Sr$_2$FeMoO$_6$ shows 
multiple features, indicating an unusual electronic state of Mo in this 
compound. In order to explore whether the multiple feature within the main 
peak region of Mo 3$d$ spectrum is a consequence of a negative $U_{eff}$, we
have performed a core level calculation considering a cluster with four 
Mo atoms arranged at the corners of a square. Each atom had 4 orbitals,
with a crystal field splitting of 0.9~eV between a triply degenerate set
($t_{2g}$-like) and a single $e_g$-like orbital.
Hopping between $t_{2g}$ orbitals was taken to be -0.4~eV, while that between
the $e_g$-like orbitals was taken to be -0.5~eV. $U_{eff}$ was
taken to be -0.5~eV. The presence of the core-hole was taken into account by
a decrease in the onsite energies by the core-hole potential of 0.9~eV \cite{uhigh}.
The resulting calculated spectrum is shown in the main figure as a solid line
for comparison. The agreement between the experimental and the calculated 
spectra is remarkably good. It was also found that the calculated spectrum 
had invariably the simple double peak structure of Mo and MoO$_3$ for any 
parameter set as long as the $U_{eff}$ was taken to be positive, 
indicating that the $U_{eff}$ at the Mo sites in indeed negative in this system,
driving an electronic phase separation.

Combining {\it ab-initio} and model Hamiltonian calculations, we 
establish the presence of strongly enhanced $I_{eff}$ and $J$,
responsible for the unusual electronic and magnetic properties
of Sr$_2$FeMoO$_6$. The enhanced $I$ results in an effective
negative $U$ at the Mo site which results in a complex core level
spectrum observed in our photoemission experiments.

We thank O.K. Andersen, M. Avignon, I. Dasgupta, 
H.R.~Krishnamurthy and B. Sriram Shastry
for useful discussions. We also thank O. Jepsen and O.K. Andersen
for providing us with the LMTO code. This research is funded by Department of
Science and Technology.

\section{figure captions}

Fig.~1. LMTO band dispersions for Sr$_2$FeMoO$_6$ along symmetry directions (see text). 

Fig.~2. The dispersions of the bands with $t_{2g}$ symmetry at the Mo 
site for (a) up spin and (b) down spin. All other
degrees of freedom have been integrated out as discussed in the text.

Fig.~3. Energy level diagram expected at the Fe (left panel) and Mo (central panel) 
sites from an ionic picture. The energy levels at Mo are modified in the presence of
Fe-Mo hopping interactions as shown in the right panel.

Fig.~4 The experimental (solid circles) and calculated (solid line)
Mo 3$d$ core level spectrum for Sr$_2$FeMoO$_6$. The experimental spectra (solid circles) for  
for MoO$_3$ and Mo-metal are also shown.


\begin{references}
\bibitem{uhigh} Usually the core-hole potential is taken to be 1.2 times $U$. In the
present calculations, we found that varying the core-hole potential in the range 0.8-1.2 eV
did not change the calculated spectrum significantly.
\end{references}
\end{document}